\documentclass[prb,twocolumn,showpacs]{revtex4} 
\usepackage{graphicx} 
\begin{document} 
\title{Supramolecular interactions \\in clusters of polar and 
polarizable molecules} 
  
\author{Francesca Terenziani} 
\affiliation{Dip. di Chimica GIAF 
 Universit\`{a} di Parma, I--43100 Parma, Italy;
INSTM-UdR Parma} 
\author{Anna Painelli} 
\email{anna.painelli@unipr.it} 
\affiliation{Dip. di Chimica GIAF 
 Universit\`{a} di Parma, I--43100 Parma, Italy;
INSTM-UdR Parma}
  
\date{\today} 
 
 \begin{abstract} 
We present a model for molecular materials made up of polar and 
polarizable molecular units. A simple two state model is adopted for
each molecular site and only classical intermolecular interactions are
accounted for, neglecting any intermolecular overlap. 
The complex and interesting physics driven by  interactions 
among polar and polarizable molecules 
becomes fairly transparent in the adopted model.
 Collective effects  are recognized in the large variation 
of the molecular polarity with 
supramolecular interactions, and cooperative behavior shows up  
with the appearance, in attractive lattices, 
of discontinuous charge crossovers.
The mf approximation proves fairly accurate in the description of the gs 
properties of MM, including static linear and non-linear optical
 susceptibilities,  apart from the region in the close proximity 
of the discontinuous charge crossover. 
Sizeable deviations from the excitonic description
are recognized both in the excitation spectrum and in  
 linear and non-linear optical responses. New and interesting phenomena 
 are recognized near the discontinuous charge crossover for
 non-centrosymmetric clusters, where the primary photoexcitation event
corresponds to  a multielectron transfer.

 \end{abstract} 
  
  
\maketitle   
  
\section{\label{intro}Introduction} 
 
The promise of molecular materials (MM) for advanced applications 
is impressive: 
not only {\it old} devices can be fabricated by light-weighted, flexible 
and cheap materials, but  a new generation of devices is likely to appear 
to drive  a molecular electronic and/or photonic revolution.  
Large non-linear responses to external perturbations (including electrical 
and optical fields) are required for applications and functional MM are 
often based 
on $\pi$-conjugated molecules and polymers. 
Structure-properties relationships are well understood at the molecular 
level at least for a few families of molecules, \cite{marder,albota} and
the guided synthesis of molecules with specified behavior has already 
been demonstrated. \cite{bd,abbotto} But as the synthetic ability evolves 
from the molecular to the supramolecular level, \cite{supra}
 interpretative tools are required to guide the synthesis at all stages, 
and supramolecular structure-properties relationships must be devised.

Extending our comprehension of  the properties of MM 
from the molecular to the supramolecular level is a challenging task since 
non-additive, collective behavior appears in MM as a results of intermolecular 
interactions. Particularly important collective effects are expected  
in functional MM  
based on large (and largely polarizable) $\pi$-conjugated molecules,  
as it was suggested by McConnell, 
40 years ago, \cite{mcconnell} 
and as it was more recently underlined with specific reference 
to MM for non-linear optical (NLO) applications.\cite{dirk}

The mean-field (mf) approach  is a simple approximation scheme to
describe the ground state (gs) of MM. Within mf the  problem of $N$ interacting
molecules factorizes  into $N$ effective molecular problems and collectivity 
is introduced by the local self-consistent  
fields generated by the surrounding at each molecular site. Recently  
the mf approximation was implemented into a quantum chemical calculation 
of the  polarization of crystals and films of large conjugated 
molecules.\cite{ts,ts1,ts2}  
The charge distribution on  
polarizable molecular sites is strongly affected by local electric fields,
and prominent collective  behavior is recognized 
 in the very large on-site charge redistribution in crystals  
of quadrupolar molecules.\cite{ts,ts1} 
So far the approach was not applied to  
crystals or films of polar molecules, where even larger effects are expected.

The mf approximation does not apply to excited states, that 
in MM are usually described within the excitonic approach,
a very popular approximation scheme originally developed 
 for weakly interacting MM.\cite{davidov,agranovich} 
In this limit sizeable collective phenomena are expected only if 
intermolecular interactions  mix up  degenerate (or quasi-degenerate) states.  
Within the excitonic model, then,
 any interaction  among  states with a different
number of excitations is disregarded. 
The mf approximation supports collective behavior, but suppresses any 
correlation of the electronic motion on different molecules.
The excitonic approximation instead describes  {\it collective and correlated} 
excited states.\cite{knoster_varenna}
 But, not accounting for the mixing of states 
with a different number of excitations, 
the excitonic model disregards the molecular polarizability, a serious  
limitation for MM of interest for advanced applications.
 Moreover, the excitonic model relies on  
the definition of molecular (local) ground and excited states,  as states 
relevant to the molecule embedded in the 
material, but no hint is given on how to  derive these effective states 
from the molecular  and supramolecular 
structure.\cite{davidov,agranovich,knoster_varenna} 

In clusters of non-polar molecules leading supramolecular interactions  
are represented by electrostatic interactions between transition 
dipole moments:
the resulting $J$-term in the excitonic model describes exciton 
hopping between different sites.\cite{agranovich}
The same interaction is also responsible  
for the so-called non-Heitler-London term:\cite{agranovich}
 this term, neglected in the 
excitonic approximation,
 mixes up states whose exciton number differs by two units.
 A recent detailed study has demonstrated that this term has 
indeed only minor effects in aggregates of non-polar 
chromophores.\cite{knoster}

The simple excitonic model derived for 
 materials made up of non-polar molecules 
is often adopted, with no further scrutiny,  to describe materials 
based on  polar molecules. In this case, however,  
additional  interactions appear due to the  
finite permanent dipole moments of the molecular units in both the ground 
and excited states. In particular, the interaction  
between permanent dipole moments on different molecular units 
introduces in the Hamiltonian an 
exciton-exciton interaction  term. This term conserves the exciton number
and therefore survives in the excitonic approximation.  
The role of the exciton-exciton interaction 
in promoting bound-exciton states (and specifically  biexcitons)
in  layers of polar molecules, was
investigated  in ref. \onlinecite{spano}. The formation
of exciton strings was also discussed in ref. \onlinecite{ezaki}
 as a consequence of exciton-exciton
interactions in models for linear aggregates. This same concept was applied
in a different context, to explain anomalous spectral features 
in differential transmission spectra of organic 
  charge-transfer (CT) salts  with a  
mixed donor-acceptor (D-A) stack motif, \cite{mazumdar_science,mazumdar}
leading to the experimental confirmation of bound exciton states.
 
The excitonic model was developed to describe MM in the limit of weak
interactions, i.e. for intermolecular energies much
smaller than  intramolecular excitation energies.\cite{davidov,agranovich}
Electrostatic interactions among polar molecules are instead large, and
can easily be of similar magnitude or even larger than the 
small excitation energies typical of largely polarizable molecules.
The failure of the excitonic model to describe interacting polar
and polarizable molecules is therefore hardly surprising.
However, since current understanding of
optical properties of MM  generally relies
 on the excitonic picture, it is important to explicitly discuss
the limits of this approximation. 
We will show that the properties of MM based on polar and polarizable 
molecules are quantitatively as well 
as qualitatively different from the sum of the molecular properties.
New and interesting physics appears in these materials
due to supramolecular interactions, and it cannot be  appreciated within 
standard approximation schemes.
 
In this paper we discuss the gs and the excitation spectrum of
 a simple model for clusters of polar and polarizable 
chromophores. Each molecular unit is described in terms of  
the Mulliken model.\cite{mulli} 
 This  two state model was originally proposed to describe
 D-A CT complexes in solution.\cite{mulli}
Later it was  adopted to model NLO responses of 
push-pull chromophores,\cite{oc}
 an interesting class of  $\pi$-conjugated molecules 
with electron D and A end groups.
  In this context the Mulliken model, extended  
to account 
for the interaction with internal vibrations and/or with solvation  
degrees of freedom, 
offered an accurate description of low-energy spectral properties of solvated  
push-pull chromophores.\cite{como,baba}
Intermolecular interactions are introduced in terms of 
classical electrostatic forces, any intermolecular 
charge resonance being disregarded (zero intermolecular overlap 
approximation).\cite{ts}
The resulting model quite naturally applies to  
MM based on push-pull chromophores. 
By choosing different geometries for the cluster,  
the model describes different 
kinds of materials, like, e.g.,  aggregates, Langmuir-Blodgett (LB) films,  
functionalized  polymers, and molecular crystals.  
 
In a different  context, one-dimensional (1D) models for  
aligned polar and polarizable units, 
only interacting via electrostatic forces  
offered a first description of the neutral-ionic (N-I) phase 
transition observed in CT crystals with a mixed stack motif.\cite{soos}  
This model is fairly crude, since it reduces a stack of overlapping  
 molecules (...D-A-D-A....)  to a collection of non-overlapping 
dimers (...D-A D-A....). 
In spite of that, a mf description of the relevant gs  
offered a first picture for the N-I transition, and 
specifically for the crossover from a continuous N-I interface to a  
discontinuous transition at large electrostatic interactions.\cite{soos} 
More recently the same model, within the excitonic approximation,  
yielded an interesting picture 
for the excitation spectrum of mixed stack CT salts with a 
largely N gs.\cite{mazumdar_science,mazumdar}

\section{\label{model}The model} 
 
The Mulliken model \cite{mulli} was originally proposed to  describe 
charge resonance in isolated D-A complexes  on the basis of the two 
 limiting neutral (N) and ionic (I) states: 
 $|DA\rangle$ and $|D^+A^-\rangle$, respectively. 
The same model describes intramolecular charge resonance in
 push-pull chromophores where the 
same two basis states describe the limiting apolar and charge-separated
(zwitterionic) structure,\cite{oc}  as schematically  
shown in Fig. 1 for a typical chromophore. Following Mulliken,  
we disregard all matrix elements of the dipole moment operator,
except  the dipole moment of the charge separated, ionic (I) state,
$\langle D^+A^-|\hat \mu |D^+A^-\rangle =ea$, where $a$ measures
the distance between D and A molecules in CT complexes or an 
effective molecular length in push-pull chromophores.
If  $2z_0$ is the energy gap between the two basis states,  
and $-\sqrt{2}t$ is the mixing matrix element (in the following
we will fix $\sqrt{2}t$  as the energy unit), 
the Mulliken Hamiltonian is written as: 
 
\begin{equation} 
H=2z_0\hat \rho - \sqrt{2}t \hat \sigma_x 
\label{hdim} 
\end{equation} 
where $\hat \rho= \hat \sigma_z +1/2$ is the ionicity operator, measuring  
the degree of CT from $D$ to $A$, and $\hat \sigma _{x/z} $ are the Pauli  
matrices.  
 
The solution of this Hamiltonian is trivial. The resulting ground  
 and excited state are linear combinations of the two basis states, and 
the amount of mixing is  defined in terms of a single parameter:

\begin{equation} 
\rho= \frac{1}{2}-\frac{z_0}{2\sqrt{z_0^2+1}}, 
\label{rhogs}
\end{equation}
as follows: 
 
\begin{eqnarray} 
|G\rangle &=& \sqrt{1-\rho} |DA\rangle +\sqrt{\rho} |D^+A^-\rangle \nonumber\\ 
|E\rangle &=& \sqrt{\rho} |DA\rangle -\sqrt{1-\rho} |D^+A^-\rangle  
\label{ge} 
\end{eqnarray} 
$\rho $ is the gs expectation value of the charge operator  
and is proportional to the gs molecular dipole moment:  
$\langle G|\hat \mu |G\rangle= ea\rho$. 
 
\begin{figure}
\begin{center}
\includegraphics* [scale=0.9] {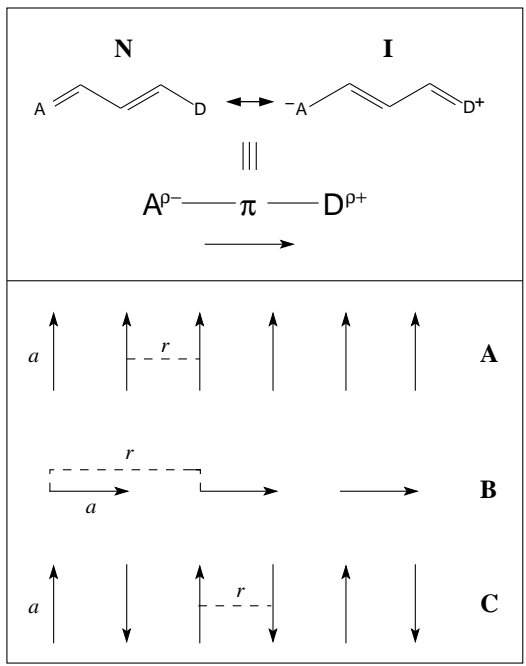}
\end{center}
\caption{Upper panel: schematic representation of a push-pull
chromophores in terms of the two limiting neutral (N) and
charge-separated, ionic (I) forms. The average charge separation in the 
ground state, $\rho$, is proportional to the molecular dipole 
moment, shown  in the figure as an arrow. Lower panel: the three 
one-dimensional lattices considered
in this paper. Each molecule is represented by an arrow, showing 
its ground-state dipole moment.}
\label{frecce}
\end{figure}We describe a MM made up of polar and polarizable molecules as a collection of 
Mulliken-molecules (i.e. of molecules described by the Hamiltonian  
in Eq.~(\ref{hdim})) interacting via classical electrostatic forces. 
The relevant Hamiltonian reads: 
 
\begin{equation} 
H=\sum_i \left( 2z_0\hat \rho_i - \sqrt{2}t \hat \sigma_{x,i} \right) 
+\sum_{i>j} V_{ij}\hat \rho_i \hat \rho_j  
\label{hami} 
\end{equation} 
where $i,j$ count the molecular sites, and the index on $\hat \rho$ and  
$\hat \sigma _x$ specifies them as  operators working on the corresponding 
 site. 
 $V_{ij}$ measures the electrostatic interaction between two  
fully-I molecules on sites $i$ and $j$.  
Different choices are possible for $V_{ij}$: dipolar interactions are  
unrealistic for typical push-pull chromophores whose length ($\sim$ 10 \AA) 
is similar to the intermolecular distances. We therefore model  
each molecule in the I ($|D^+A^-\rangle$) state
 as a segment of length $a$ bearing $+/-$ 
charges at the $D/A$ ends, and then calculate $V_{ij}$ based 
on unscreened Coulomb interactions. Specifically, for  the three
1D lattices sketched in Fig. \ref{frecce} we have:

\begin{equation}
V_{ij}=(\pm 1)^{i-j}2vw\left[\frac{1}{d_{ij}}-
\frac{1}{\sqrt{d_{ij}^2+w^2}}\right]
\end{equation}
where the $+/-$ sign refers to A/C lattices, and

\begin{equation}
V_{ij}=2vw\left[\frac{1}{d_{ij}}-
\frac{d_{ij}}{d_{ij}^2-w^2}\right]
\end{equation}
for B lattices.
In all cases $d_{ij}$ is the distance between sites $i$ and $j$,
in units of the distance $r$ between adjacent sites, $v=e^2/a$ is the 
magnitude of the electrostatic interaction between two charges at 
distance $a$, and $w=a/r$. In the following we consider 
 $v=$1 or 2, that, for typical $\sqrt{2}t=$ 1 eV values,\cite{como,baba} 
correspond to molecular lengths $a \sim 14$ or 7 \AA, respectively.
Results will be presented for $w$ increasing from 0 (the gas phase limit)
up to a few units (intermolecular distance 
corresponding to a fraction of the molecular length). For B-lattices
$w=1$ represents an upper bound.
For clusters of $N$ molecules the above  Hamiltonian is written on the  
$2^N$ basis obtained from the direct product of the two basis states located 
on each molecular site.
The eigenstates 
are obtained by exact diagonalization 
for systems with up to 16 site, by exploiting translational symmetry.

\section{\label{appro}the mean field and excitonic approaches}
In order to better appreciate the physics of 
interacting polar and polarizable molecules, we will 
compare exact solutions of the Hamiltonian in Eq.~(\ref{hami})
with those obtained within the mf and excitonic approximations. 
Both approximations  are based on the assumption that in the  
gs the electronic motion on different sites is  uncorrelated, so that 
the gs  describes a collection of  molecules each  one in the 
local (molecular) gs: $|0\rangle =|G_1 G_2....G_i....\rangle$, where  
$|0\rangle$ represents the mf gs wavefunction for the cluster, and  
$|G_i\rangle$ is the gs wavefunction for the $i$-th site.  
For a cluster of non-overlapping 
Mulliken molecules the local gs, $|G_i\rangle$,
 necessarily retains 
the same form as  in Eq.~(\ref{ge}), but with $\rho$ depending on the 
surrounding.
In any case, for any given $\rho$, $|G_i\rangle$ and  
$|E_i\rangle$ define a couple of local mutually orthogonal  
wavefunctions, and the  
 $2^N$ basis functions obtained from the direct product of these  local  
states is a complete basis for the cluster.
We define the Pauli matrix operators working on the 
local $|G_i\rangle$ and $|E_i\rangle$ basis, $\hat S_{z/x,i}$, as 
linear combinations of the Pauli operators,  $\hat \sigma_{z/x,i}$,
 working on the original $|DA\rangle$ and $|D^+A^-\rangle$ basis, as follows:
 
\begin{eqnarray} 
\hat S_{z,i}&=& (1-2\rho)\hat \sigma_{z,i}  
+2\sqrt{\rho(1-\rho)}\hat \sigma_{x,i} \nonumber\\  
\hat S_{x,i}&=& - 2\sqrt{\rho(1-\rho)} \hat \sigma_{z,i}  
+(1-2\rho)\hat \sigma_{x,i}  
\label{rotation} 
\end{eqnarray} 
To make contact with standard notation\cite{agranovich} we construct
 the following  hard-core boson  operators: 
 
\begin{eqnarray} 
\hat S_{x,i} \nonumber & = & (\hat b^\dagger_i + \hat b_i )\\ 
\hat S_{z,i}&=&1-2\hat  b^\dagger_i \hat b_i 
\label{bb} 
\end{eqnarray}
These operators have a very clear physical meaning: 
working on the vacuum state $|0\rangle$, defined as the state where
all sites are in the local $|G_i\rangle $ state, 
 $\hat b_i^\dagger$ switches the $i$-th site to the 
local  $|E_i\rangle$ state, and $\hat n_i= \hat  b^\dagger_i \hat b_i$ 
returns the 0 and 1 value for the $|G_i\rangle$ and $|E_i\rangle$ state, 
respectively.
In terms of these operators the  Hamiltonian in Eq.~(\ref{hami}) reads: 
 
\begin{equation} 
H=H_{mf}+H_{ex}+H_{uex} 
\label{htot} 
\end{equation} 
where: 
\begin{eqnarray} 
H_{mf}&=& \sum_i\left[2(1-2\rho)(z_0+M\rho) 
          +4\sqrt{\rho(1-\rho)}\right]\hat n_i \label{hmf} \\ 
&& +\sum_i\left[ 2\sqrt{\rho(1-\rho)}(z_0+M\rho)-(1-2\rho)\right]  
(\hat b_i^\dagger +\hat b_i) \nonumber \\ 
H_{ex}&=& \sum_{i>j} V_{ij}\left[\rho(1-\rho) (\hat b_i^\dagger \hat b_j 
+\hat b_j \hat b_i ^\dagger) +(1-2\rho)^2 \hat n_i \hat n_j \right] 
\label{hex}\\  
H_{uex} &=& \sum_{i>j} V_{ij}\left[\rho(1-\rho) (\hat b_i \hat b_j 
+\hat b_j ^\dagger \hat b_i ^\dagger) \right.\nonumber\\ 
&&\left.+2 (1-2\rho) \sqrt{\rho(1-\rho)} 
 (\hat b_i^\dagger +\hat b_i)  \hat n_j \right] \label{huex} 
\end{eqnarray} 
and $M=\sum_{i>j} V_{ij}/N$.

Three terms in the Hamiltonian in Eq.(\ref{htot}) 
have been conveniently grouped: 
the first one, $H_{mf}$, collects on-site terms, and, as it will be discussed 
below,  sets the basis for the mf description of the gs. Both $H_{ex}$ and
 $H_{uex}$ work on two sites: $H_{ex}$ conserves the total number of
 excitations and hence enters the excitonic Hamiltonian, whereas $H_{uex}$ 
mixes up states differing by one or two excitation, it  represents an 
ultraexcitonic term that will be disregarded in the excitonic approximation.

The transformation in Eq.~(\ref{rotation}) holds true for any $0\le\rho\le1$,
and the total Hamiltonian in Eq.~(\ref{htot}) is not affected by the specific
choice of $\rho$. Of course, its partitioning into the three terms instead
varies with $\rho$, and different excitonic approximation can be derived
from the same model Hamiltonian, depending on the specific choice of $\rho$, 
or, equivalently, on the specific choice of the local $|G_i\rangle$ and 
$|E_i\rangle$ wavefunctions. It is important to realize, however,
 that the second term in 
$H_{mf}$ does not conserve the exciton number and must be killed in the
excitonic approximation. We therefore define the {\it best}
 excitonic Hamiltonian
for our model by fixing $\rho$ as to impose the exact vanishing of
this term:
\begin{equation}
2\sqrt{\rho(1-\rho)}(z_0+M\rho)-(1-2\rho)=0
\label{rhomf}
\end{equation}
This equation defines the best local
basis for the relevant Hamiltonian in terms of the  same $|G_i\rangle$ 
and $|E_i\rangle$ states  defined 
 for the isolated molecule in Eq.~(\ref{ge}), but with  $z=z_0+M\rho$ 
playing the same role as $z_0$ in the definition of $\rho$ in 
 Eq~(\ref{rhomf}). With this specific choice of $\rho$ the 
uncorrelated gs $|0\rangle=
|G_1G_2...G_i....\rangle$ coincides with  the gs  of the 
excitonic Hamiltonian, i.e. of $H_{mf}+H_{exc}$. Since $|0\rangle$ is
the gs of an Hamiltonian where all on-site terms are retained,
it also  represents {\it the best uncorrelated gs} for the system and 
indeed coincides with the mf gs. Within mf   the cluster is 
modeled as a collection of non-interacting
molecules,  each molecules experiencing the electric field generated 
by the surrounding.\cite{ts,soos} This local electric field is  responsible 
for the renormalization of the energy gap between the local I and N 
states from  the value relevant 
to the isolated  molecule ($2z_0$)  to  $2(z_0+M\rho)$, 
relevant to the molecule
embedded in a cluster of molecules with average polarity $\rho$.\cite{soos}

The solution of the two-state problem defined by the local mf 
Hamiltonian fully defines the excitonic and ultraexcitonic Hamiltonians.
 Exciting a molecule costs an energy
$\hbar \omega_{CT}=1/\sqrt{\rho (1-\rho)}$, and, with $\rho$ fixed at
the mf value, $H_{mf}$ can be rewritten as:
\begin{equation}
H_{mf}= \hbar \omega_{CT}\sum_i n_i 
\label{hmfsimple}
\end{equation}
This  Hamiltonian  counts the number of excited molecules in the cluster
 and assigns energy $\hbar \omega_{CT}$ to each of them.  
We can also write explicit expressions for the local transition dipole moment,
$\mu_{CT}= \mu_0 \sqrt{\rho (1-\rho)}$, and for  the local mesomeric dipole
 moment (the difference between the dipole moment in the $|E_i\rangle$ and
$|G_i\rangle$), $\Delta\mu =  \mu_0 (1-2\rho)$.
Then, the first term in $H_{ex}$ (Eq.~(\ref{hex})) describes
the interactions between transition dipole moments on different
 molecules: it corresponds to  $J$-type interactions  in the standard 
excitonic  model, the only term, beyond 
$H_{mf}$ in Eq.~(\ref{hmfsimple}), that survives in the excitonic 
approximation for apolar molecules.\cite{agranovich}
 The second term in Eq.~(\ref{hex})
 accounts for electrostatic interactions between 
permanent molecular dipole moments: it vanishes for apolar molecules. 
$H_{uex}$ in Eq.~(\ref{huex}) collects ultraexcitonic terms, i.e. terms 
that mix up states with a different number of excitons. The first term 
in Eq.~(\ref{huex}) has the same origin as the exciton hopping, and is 
usually referred to as the non Heitler-London term.\cite{knoster} 
The second term in the ultraexcitonic Hamiltonian vanishes for apolar
 molecules.

Of course the Hamiltonian in Eq.~(\ref{htot}) exactly maps on the general  
Hamiltonian derived  by Agranovich for a  cluster of two-level molecules
only interacting via electrostatic forces.\cite{agranovich}
However, in the standard excitonic approximation to the general Hamiltonian,  
the local ground and excited states coincide with the eigenstates of 
the isolated molecule.
Environmental contributions to the excitation energies (the term 
proportional to $M\rho$ in the first term of $H_{mf}$) are accounted
for, but any mixing of the two local eigenstates due to environmental effect 
is neglected (the term proportional to $M\rho$ in the second term in $H_{mf}$
is disregarded).\cite{agranovich}
This amounts to a complete neglect of the molecular polarizability:
the interaction with the surrounding only  affects excitation energies, 
but not local wavefunctions. 
Most often the local ground and excited states are defined in terms of
 effective molecular states as affected by environmental interactions, 
but the parameters entering the excitonic model are then freely adjustable
and can hardly be  related to the molecular structure and/or to the 
supramolecular arrangement.
Our choice of the local ground and excited states as the eigenstates
of $H_{mf}$ not only defines the best excitonic
basis and relates the excitonic and mf pictures, but also leads to
an unambiguous definition of the excitonic and ultraexcitonic Hamiltonians. 
In fact, the mf model relevant to a
 cluster of non-overlapping molecules is
 defined, for any specific cluster geometry,
  given the model  Hamiltonian for the isolated molecule.\cite{ts} 
At variance with the standard implementation of the excitonic model,
for either apolar or polar molecules, in our approach all the 
parameters entering the excitonic (and ultraexcitonic) Hamiltonian, 
including the excitation energy, 
the exciton hopping term and the exciton-exciton interaction,
are defined by the solution of the (self-consistent)
mf local Hamiltonian and can be related to the molecular 
and supramolecular structure. 
The solution of the local mf problem is a trivial task for clusters of 
polar and polarizable molecules described in this paper, where only two
states are enough to capture the (low-energy) molecular physics. 
For MM, like  those made up of non-polar molecules, where 
dispersion forces dominate intermolecular interactions, setting up and 
solving the local mf problem is a more delicate problem since several 
molecular states must be explicitely accounted for.
In any case, uncovering  the relation between the local
mf solution of the cluster Hamiltonian and its excitonic (and ultraexcitonic)
description is an important result since it allows to set up the
description of a MM based on  information relevant to the isolated
molecule, opening the way to understand the properties of MM from
the molecular up to the supramolecular level.

\section {\label{gs}The ground state}

\begin{figure}
\begin{center}
\includegraphics* [scale=0.5] {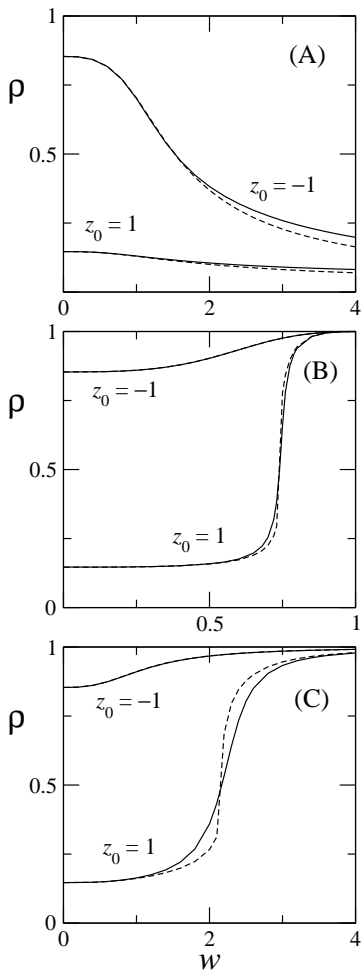}
\end{center}
\caption{Ionicity vs the inverse intermolecular distance for 
A, B and C, lattices with $v$=1 and $z_0=\pm 1$ 
Dashed lines: mf results for $N=\infty$; 
continuous lines: exact results for $N=16$.}
\label{rho_cont}
\end{figure}

Figure \ref{rho_cont} shows the evolution of the gs polarity vs the inverse 
intermolecular distance for  A, B, and C clusters
with $v=1$ and two different $z_0$ values. In cluster A
 repulsive intermolecular interactions disfavor the charge separation 
on molecular sites and $\rho$ decreases with $w$. Just the opposite
 occurs in B and C clusters where attractive interactions favor charge
 separation. An isolated molecule with an almost neutral gs 
($\rho \sim $ 0.15 at $w=0$ for $z_0=1$)  becomes even more neutral in the
repulsive, A-cluster, but, with increasing intermolecular interactions,
it becomes more and more I in attractive (B or C) lattices. 
For the data in Fig.\ref{rho_cont} , the N chromophore crosses the $\rho=0.5$ 
interface separating a N from an I gs at $w \sim 2$ in C-lattice, i.e. at an
inter-chromophore separation $\sim 5$ \AA~ for typical molecular length $\sim
10 $ \AA. In B-lattices the interface is located at $w \sim 0.75$, 
corresponding to intermolecular contacts $\sim $ 3 \AA.
Conversely, molecules with a zwitterionic gs in the gas phase
($\rho \sim$ 0.85 at $w=0$ for $z_0$=-1) can be driven to a N gs
in the repulsive, A-lattice 
at $w \sim 1.5$, corresponding to intermolecular distances $\sim$
 7 \AA.    Even more interesting is 
the observation that the same molecule in different supramolecular 
arrangements can have  qualitatively  
different gs, and hence distinctively different properties. Consider e.g. 
an isolated molecule with an N gs ($z_0=1$): in an A-type lattice
with  $w\sim 3$ (i.e. for  intermolecular
 distances $\sim$ one third of the molecular length)
the molecular polarity stays basically unaffected (it slightly 
decreases down to $\sim$ 0.1). The same molecule in a C-lattice with similar 
intermolecular distance becomes instead  very  polar, with $\rho \sim$ 
0.9 corresponding to an almost zwitterionic gs. 
Since the properties of push-pull chromophores largely 
depend of $\rho$,\cite{apcpl}
very different behavior is predicted for different MM made up by
the same molecular unit.

\begin{figure}
\begin{center}
\includegraphics* [scale=0.5] {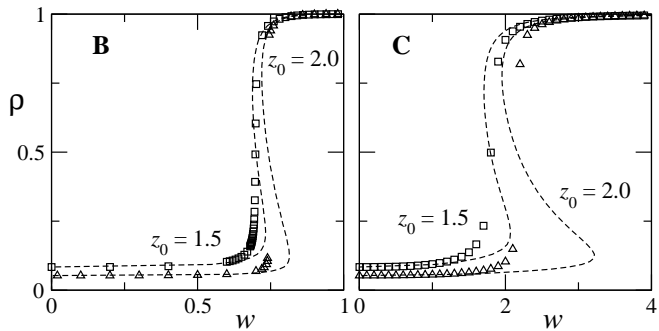}
\end{center}
\caption{Exact ionicity vs the inverse intermolecular distance for 
attractive, B and C, lattices with $N$= 16, $v$=2 and $z_0$= 1.5 (squares)
and 2 (circles). Dashed line show $N=\infty$ mf results.}
\label{rho_disc}
\end{figure}

Dashed lines in Fig  \ref{rho_cont}
show mf results for the gs polarity: as far as 
$\rho$ is concerned, the mf approach
 works well at least for not too large interactions. 
Within mf we can also understand the qualitatively different
behavior of attractive and repulsive lattices. 
As discussed in the previous Section, within mf  $\rho$ is
given by Eq.~(\ref{rhogs}), but with $z=z_0+M\rho$ playing the same role 
of $z_0$, so that:\cite{soos}
\begin{equation}
\frac{\partial \rho}{\partial z_0}= 
\frac{d \rho}{dz}\left(1-M \frac{d \rho}{dz}\right)^{-1}
\end{equation}
$\rho(z)$ has a negative slope, so that 
 for repulsive lattices with  $M>0$, the  slope of  $ \rho( z_0)$ 
decreases in magnitude with $M$, justifying the smooth evolution of the 
$\rho(w)$ curves in Fig.\ref{rho_cont}A. Just the opposite occurs for 
attractive lattices ($M<0$). In this case
 $\partial \rho/\partial z_0$ becomes more negative with increasing
the strength of inter-site interactions and,  for $M <-2$, it changes 
its sign, marking the occurrence 
of a discontinuous phase transition from the N to the I regime. 
The mf N-I crossover is located  at $z =0$, and  S-shaped $\rho(w)$ curves are
therefore calculated within mf for systems with   $z_0 > 1$, as shown in 
Fig. \ref{rho_disc}
 for B and C lattices with $v=2$ and  $z_0=1.5$ and 2.
Regions with negative slope in the mf $\rho(w)$ curves 
correspond to unstable states, and a bistable regime appears where two stable
states ($\partial \rho/\partial w >0$) coexist for the same $w$.\cite{soos}
Demonstrating the appearance of 
a discontinuous interface from exact diagonalization 
of finite clusters Hamiltonians is hardly possible. In fact
 only {\it stable} states are accessible by exact diagonalization: 
neither the unstable
nor metastable states appear in exact curves. In any case,
results from exact diagonalization (symbols
 in Fig. \ref{rho_disc}) show, at large $z_0$,
 a well pronounced discontinuity, 
 consistent with the discontinuous behavior recognized within mf.

Isolated (gas phase or solvated) molecules cannot support any kind of 
phase transitions: the appearance of a discontinuous N to I crossover is 
one of the most impressive evidences 
 of the cooperative behavior of MM made up of polar
and polarizable chromophores.
In the proximity of the discontinuous crossover large deviations from 
the mf picture are expected on general grounds, due to the failure 
of the non-correlated description of the gs. 
Correlations effects in this region 
have important spectroscopic consequences
and  will be discussed in detail in Sect. \ref{disco}. 

\section{\label{spectrum}The excitation spectrum}

\begin{figure}
\begin{center}
\includegraphics* [scale=0.7] {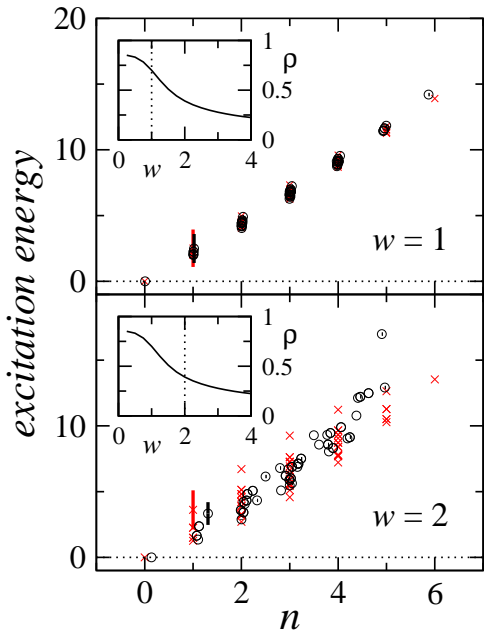}
\end{center}
\caption{Excitation spectrum of an A-lattice with $N=6$,
 $v$=1, $z_0=-1$ and two different $w$.
Circles and crosses show the  excitation energy for 
exact and mf eigenstates, respectively, against the 
number of excitons.
States on the zero energy axis correspond
to the gs.  Error bars measure the squared transition 
dipole moment from the gs to the relevant states.
Insets show the $\rho(w)$ mf curve for the relevant parameters,
with the dotted vertical line marking the $w$ value for which results are
reported in the parent panel.}
\label{eccia}
\end{figure}

\begin{figure}
\begin{center}
\includegraphics* [scale=0.7] {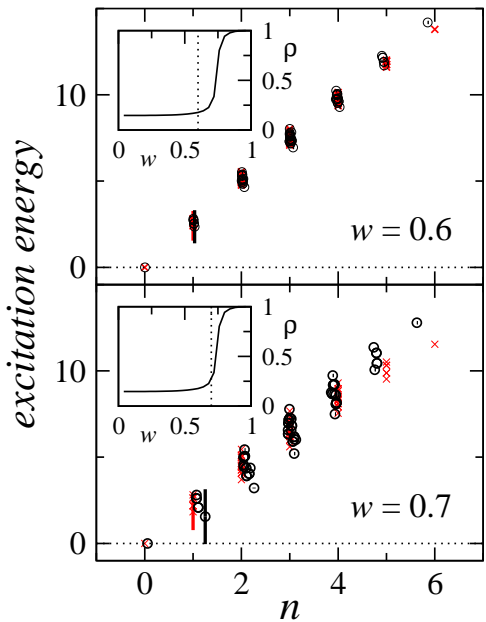}
\end{center}
\caption{The same as in Fig. \ref{eccia}, but for B-lattices
with $v=1$, $z_0$=1 and two different $w$.}
\label{eccib}
\end{figure}

\begin{figure}
\begin{center}
\includegraphics* [scale=0.7] {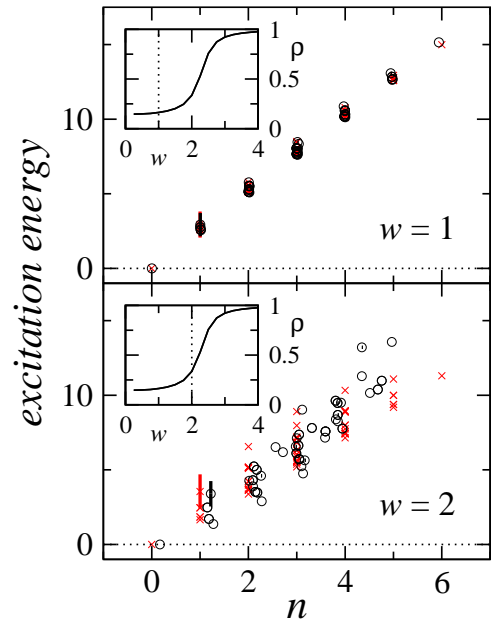}
\end{center}
\caption{The same as in Fig. \ref{eccia}, but for C-lattices
with $v=1$, $z_0$=1 and two different $w$.}
\label{eccic}
\end{figure}

The solution of the local mf problem defines the best local  basis 
$|G_i\rangle$ and $|E_i\rangle$ for the excitonic problem and hence the optimal
excitonic approximation for the Hamiltonian in Eq.~(\ref{hami}).
Figs. \ref{eccia}, \ref{eccib}, and \ref{eccic},  compare
 the complete set of exact eigenstates
 (circles) with the excitonic eigenstates (crosses) for A, B, and 
C clusters, respectively. 
To avoid overcrowding, data are shown for $N=6$.
In the figures, the excitation energy 
(i.e. the energy minus the gs energy) is reported 
for each eigenstate vs $n=\langle \sum_i \hat n_i \rangle $, 
the exciton number. We set  $z_0=-/+ 1$ for
repulsive/attractive lattices (cf Fig. \ref{rho_cont}). 
For each cluster we consider
 two different $w$ values, corresponding to interactions 
of weak and medium strength (upper and lower panel, respectively).
For reference purposes, the inset in each  panel
 shows the mf  $\rho(w)$ curve, with the vertical dotted line marking
the relevant $w$ values.

Within the excitonic approximation $n$ is conserved: excitonic eigenstates 
line up exactly in Figs.  \ref{eccia}, \ref{eccib}, and \ref{eccic},
 defining bands of states with integer $n$.
The excitonic bandwidth measures the amount of 
 mixing of states with the same $n$, induced by 
the exciton hopping (the first term in Eq.~(\ref{hex}), 
corresponding to $J$-like interactions in the standard excitonic model) 
and by the  exciton-exciton interaction term (the second term in
 Eq.~(\ref{hex}), relevant to polar molecules). This last term affects 
the energy of multi-exciton states lowering (increasing) the energy of 
states with several nearby excitations in attractive (repulsive) lattices.

The ultraexcitonic Hamiltonian in Eq.~(\ref{huex}) mixes up states with 
different $n$. For weak interactions (upper panels in Fig. 
 \ref{eccia}, \ref{eccib}, and \ref{eccic})
the ultraexcitonic mixing is not large, and $n$ is approximately 
conserved also for the exact eigenstates (circles). 
But, for larger interactions (bottom panels), the combined role of 
exciton-exciton interactions and of the ultraexcitonic mixing can have 
important consequences. 

Figs.  \ref{eccia}, \ref{eccib}, and \ref{eccic}
 carry one more information: The size of the error-bars attached 
to each symbol are proportional to the transition dipole moment from 
the gs toward the relevant eigenstate.  A general selection rule applies 
to both exact and excitonic eigenstates:  only 
zero wavevector ($k=0$) states are reached 
by optical transition. Moreover, within the excitonic approximation,
only one excitation can be created upon one-photon absorption so that only 
the state with $k=0$ and  $n=1$  has a finite transition dipole moment
from the gs. 
This state is at the top of the $n=1$ exciton band in the repulsive A lattice 
($J>0$, H-aggregates) and  at the bottom of the 
one-exciton band in the attractive, B-lattice ($J<0$, J-aggregate). 
C-lattice with two molecules per unit cell is different: 
 the inversion symmetry that relates
the two molecules in the unit cell allows to classify  
the two $k=0$ one-exciton states
as symmetric and antisymmetric, and only the antisymmetric 
state is optically active. 
This state corresponds to a $k=\pi$ state in the extended zone 
representation, and for this {\it attractive} lattice ($J<0$) the optically 
allowed state actually lies at the top of the one-exciton band.

The ultraexcitonic mixing relaxes the $\Delta n=1$ selection rule of the
 excitonic model and the oscillator strength can spread over
 several $k=0$ states. 
However, at least for the cases shown in Figs.  
\ref{eccia}, \ref{eccib}, and \ref{eccic}, 
a  single excited state retains  most of the oscillator 
strength, leading to
a  spectral behavior qualitatively similar as in the excitonic picture. 
However, for not too weak interactions 
(bottom  panels in Figs. \ref{eccia}, \ref{eccib}, and \ref{eccic}) 
the exciton number in the most optically allowed state 
distinctively deviates from the excitonic result ($n=1$).
 More generally,
the excitonic description of the photoexcited state is 
not very accurate. Consider, just as an example, a B-lattice with parameters 
as in Fig. \ref{eccib}, upper panel. As for the data reported in the
figure are concerned, the excitonic 
picture  works fairly well. However, the average ionicity of the 
optically allowed state calculated in the excitonic picture is 
about $\sim$ 15\% smaller than the exact result. The 
difference becomes $\sim$ 40\% for the parameters
in the bottom panel of the same figure. The discrepancy between excitonic
and exact descriptions of the states reached upon photoexcitation
 becomes more and more evident 
as the discontinuous charge crossover is approached, and
 will be discussed in greater detail in Sect. \ref{disco}.

\section{\label{nlo}Static susceptibilities}

\begin{figure}
\begin{center}
\includegraphics* [scale=0.5] {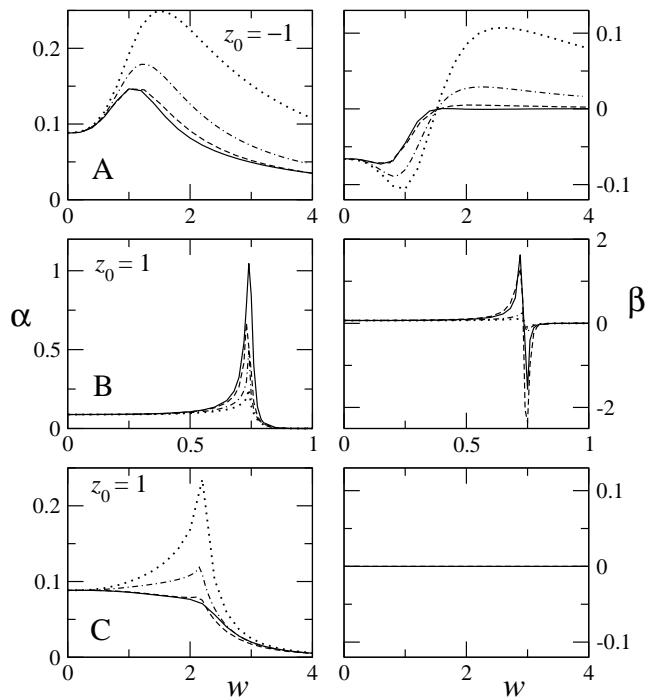}
\end{center}
\caption{Linear susceptibility ($\alpha$) and first hyperpolarizability
($\beta$) vs the inverse intermolecular distance for 
A, B and C clusters (upper, middle and lower panels, respectively) with $N=16$,
$v$=1 and $z_0=\pm 1$. Dotted lines: mf-oriented gas approximation; 
dot-dashed line: excitonic approximation; dashed lines:
mf approximation;
continuous lines: exact results. The $\beta$ response for $C$-clusters
vanishes by symmetry.}
\label{nlo_ff}
\end{figure}

Collective effects that are discussed in the previous 
Section with reference to excitation spectra,  must also  show up in linear
and non-linear susceptibilities of  MM.
Here we focus attention on the linear susceptibility $\alpha$ and on the 
first hyperpolarizability $\beta$, defined by the expansion of 
the gs polarization, $P$ (the dipole moment per unit volume) 
vs a static electric field, $F$, as follows:

\begin{equation} 
P(F)=P(0)+\alpha F +\frac{1}{2} \beta F^2+...
\label{pdif}
\end{equation}
Continuous lines in Fig. \ref{nlo_ff} report exact $\alpha$ and $\beta$ values 
calculated as a function of $w$ for 16-site clusters with  $v=1$. 
Top panels refer
to an A-cluster with $z_0=-1$, middle and bottom  panels refer to 
B an C clusters, respectively, with $z_0=1$.
The cluster susceptibilities 
deviate considerably from  those relevant to the isolated molecules,
and a  large and non-trivial dependence of the optical responses 
on the supramolecular arrangement is observed.
This is due to the large response of polarizable molecules 
to the local electric fields in the material, and has important, and so far 
not fully appreciated, consequences. 

Some degree of non-additivity of the optical responses of clusters 
of push-pull chromophores could have been guessed from the large 
dependence of the gs molecular polarity on supramolecular interactions, 
as shown in Fig. \ref{rho_cont} and \ref{rho_disc}.
 For push-pull chromophores in fact 
linear and non-linear optical responses strongly depend on $\rho$. 
Since closed expressions for $\alpha$ and $\beta$ as a function of $\rho$ 
are available,\cite{apcpl}
 the simplest approach to static susceptibilities relies on 
a {\it mf oriented gas} picture. Within  mf  the gs of the cluster 
describes  a collection of non-interacting molecules.
 The responses of a  cluster of molecules can accordingly be calculated as
the  sum of contributions from the  molecules in the relevant mf gs.
The  susceptibilities obtained in this approximation (dotted lines in Fig. 6)
deviates considerably from exact results, a particularly  impressive result 
in view of the reliability of mf estimates of $\rho$ for the same clusters 
 (cf Fig. \ref{rho_cont}). The oriented gas approach to 
susceptibilities fails since the response of a molecule to an applied
field strongly depends on the molecular environment. 

As a matter of fact, the mf approximation leads to fairly 
accurate estimates of the susceptibilities  for MM, provided 
the oriented-gas approximation is relaxed. Dashed lines in Fig. \ref{nlo_ff}
show  the susceptibilities calculated as successive $F$-derivatives of the 
polarization of the cluster, calculated in the mf approximation. 
Apart from small deviations that appear just in the crossover regime 
where the mf estimate of $\rho$ is slightly inaccurate by itself
(cf Fig. \ref{rho_disc}), 
the mf susceptibilities nicely agree with exact results. 
 The large deviations between the susceptibilities calculated as sum of 
molecular mf contributions (mf oriented gas approximation, 
dotted lines in Fig. \ref{nlo_ff}) and those obtained  from the successive 
derivatives of the mf polarization of the cluster (mf results, dashed lines
in Fig. \ref{nlo_ff}) are a direct measure of
 collective effects, and can be understood since the polarization  
of the cluster is  the sum of the  (oriented) dipole moments calculated 
for each molecule, but its derivatives
 with respect an applied field largely deviate from 
 the sum of the corresponding 
derivatives (as imposed in the oriented gas model).

Sum over state (SOS) expressions  directly
link static susceptibilities to the excitation spectrum, \cite{ow}
 offering a way to  estimate the magnitude of 
excitonic and ultraexcitonic contributions to  static optical responses. 
For sure susceptibilities evaluated from the successive derivatives of the 
gs polarization and from SOS expressions do coincide, 
provided they are obtained from  the eigenstate of the same 
Hamiltonian.\cite{jcp} 
Then exact SOS susceptibilities coincide with the $F$ derivatives 
of the exact gs polarization.
Similarly, the  susceptibilities obtained in the mf oriented gas approach,
coincide with the SOS susceptibilities calculated for a collection 
of non-interacting (mf) molecules.
The large deviations of the mf-oriented gas susceptibilities from exact 
results can then be ascribed to the combined effect of excitonic and 
ultraexcitonic mixing in the excitation spectrum.
 The complete diagonalization of the excitonic Hamiltonian ($H_{mf} +H_{ex}$)
 and the subsequent calculation of SOS susceptibilities leads to 
results reported as dot-dashed 
lines in Fig. \ref{nlo_ff}. Accounting for excitonic interactions
considerably improves over the oriented gas estimate 
 of static susceptibilities. However,  sizeable deviations from 
exact results are observed, pointing to the importance of ultra-excitonic 
effects. In particular we underline that large ultraexcitonic contributions
to both $\alpha$ and $\beta$
are observed  even in regions where the mf 
approximation gives a fairly accurate description of both the gs polarity 
and of the static susceptibilities. Neglecting the molecular polarizability
in  the description of excited states,
as imposed in the excitonic approach,
 is a dangerous approximation for  linear and 
non-linear optical responses of molecular materials made up of polar and 
polarizable molecules.

\section{\label{disco}collective and cooperative effects at the discontinuous
charge crossover}
\begin{figure}
\begin{center}
\includegraphics* [scale=0.7] {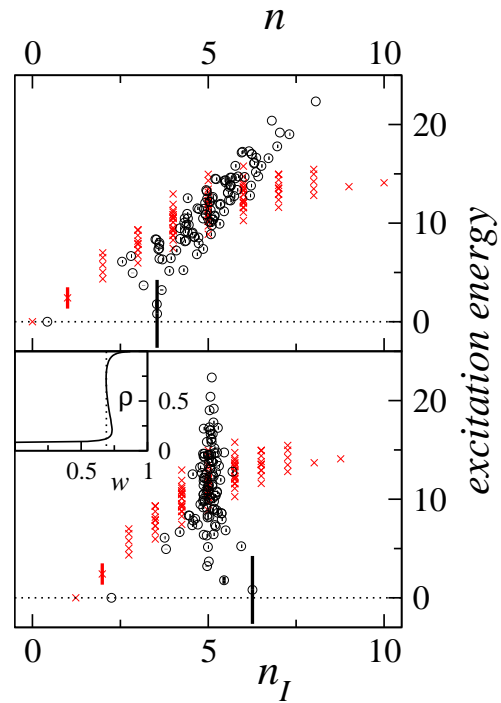}
\end{center}
\caption{Excitation energies for  the zero wavevector eigenstates of 
a 10-site B-cluster with $v=2$, $z_0=1.5$ and $w=0.69$
 reported against  the number of excitons (upper panels)
and the number of I molecules (bottom panels). Circles and
crosses refer to exact and excitonic eigenstates, respectively.
States on the zero energy axis correspond
to the gs.  Error bars measure the squared transition 
dipole moment from the gs to the relevant states. The inset shows the 
corresponding  $\rho(w)$ mf curve, with the vertical dotted
 line marking the relevant $w$ value.}
\label{ecci_disc_b}
\end{figure}

\begin{figure}
\begin{center}
\includegraphics* [scale=0.7] {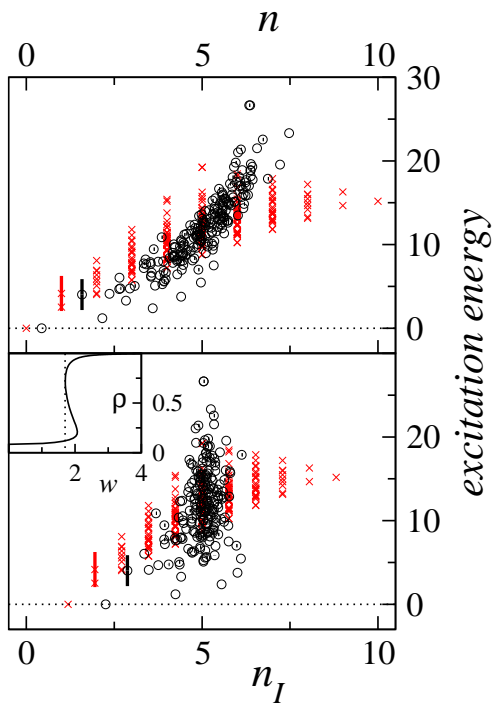}
\end{center}
\caption{The same as in Fig. \ref{ecci_disc_b} for a 
C cluster with $v=2$, $z_0=1.5$ and $w=1.7$. For graphical purposes the
error bars measuring the transition dipole moments have been magnified by
a factor of two with respect to those reported in Fig. \ref{ecci_disc_b}.}
\label{ecci_disc_c}
\end{figure}

As discussed in Sect. \ref{appro},  the mf approximation  applied to 
 attractive lattices with $z_0 >1$ 
 leads to  S-shaped $\rho(w)$ curves
(cf Fig \ref{rho_disc}), 
that signal the occurrence of a discontinuous phase transition. 
The excitation spectrum of the system near the discontinuous
charge crossover is interesting and deserves detailed discussion.
 Figs. \ref{ecci_disc_b} and  \ref{ecci_disc_c}
 show the excitation spectrum calculated for B and C clusters
 in the N regime, but very near to  the discontinuous 
crossover. Results are shown for clusters of 
10 sites and, for the sake of clarity, only zero wavevector eigenstates are
reported, with circles and crosses referring to exact and excitonic
results, respectively. 

Near the discontinuous crossover the excitation energy ($\hbar \omega_{CT}$
in Eq.~(\ref{hmfsimple}))
is at a minimum and exciton hopping and  exciton-exciton interactions 
are  large enough to induce a partial overlap of
excitonic bands (cf crosses in Figs \ref{ecci_disc_b}, \ref{ecci_disc_c}). 
Ultraexcitonic mixing is therefore very effective:
the exciton  number is not even approximately conserved in exact eigenstates.
Moreover, the energy of exact excited states show a non-monotonic
dependence on the exciton number. In particular,
 in B and C lattices 
the lowest excited state corresponds to a state  with
more than 3 and 2 excitons, respectively. This result contrasts sharply with 
the excitonic picture, where the excitation energy increases monotonously 
with the exciton number, so that  the lowest excitation energy corresponds
to the creation of a single exciton.

As  a matter of fact,  the excitonic picture is completely
spoiled in the bistability region. Whereas this is  not particularly 
surprising, it is important to realize
 that many concepts usually adopted
to discuss excitations in molecular crystals fail there. In
particular, the very same concept of 
excitation is challenged by exact results for B and C clusters near
 the discontinuous crossover.
 Both  lattices in fact have in the  gs a number of
excitons distinctively larger than zero (cf  Figs.\ref{ecci_disc_b}
and \ref{ecci_disc_c} where the exact gs eigenstate has
$n>0$). In these conditions the definition of a local excitation
 is somewhat artificial. We therefore 
 go back to the original $|DA\rangle$ and
$|D^+A^-\rangle$ basis and, in the lower panels in  Figs.\ref{ecci_disc_b}
and \ref{ecci_disc_c} we report the same excitation spectrum 
as in the upper panels, but with the abscissa axis now measuring 
the average number of fully-I ($|D^+A^-\rangle$) sites, i.e.
 the expectation value of $\hat n_I=\sum_i\hat \rho_i$. 
The operator $\hat n_I$ is diagonal on the excitonic eigenstates:
in the excitonic approximation $n_I$ and $n$ are related by a simple
expression, 
$n_I=N\rho+(1-2\rho)n$, where $n_I$, and $n$ 
are the expectation values of the corresponding 
operators in the given state and $\rho$ is the mf molecular ionicity. 
Excitonic eigenstates then 
line up vertically also when reported against $n_I$ (cf crosses in Figs.
\ref{ecci_disc_b} and \ref{ecci_disc_c}).
For exact eigenstates instead  $n$ and $n_I$ carry independent information.
 
In any case, $n_I$ is finite in  both the exact and excitonic gs
 to indicate a finite $\rho$.
 The difference between the excitonic and exact $n_I$ 
of course parallels the sizeable  difference between the  mf
and exact $\rho$ observed  near the discontinuous crossover 
(cf Fig. \ref{rho_disc}).  More interesting is the behavior of
the lowest excited state: in both B and C lattices the lowest
excitation corresponds to a state whose  $n_I$ appreciably differs
 from the gs value: the lowest exact 
 excited states has a very different nature from the gs. This contrasts
with the excitonic picture where the lowest excited state is a 
state with a single excitation,
 whose ionicity differs from the gs ionicity by $|1-2\rho|\le$1.

Attractive supramolecular interactions in B and C lattices lead
to a qualitatively similar  $\rho(w)$ dependence  (cf Figs.\ref{rho_cont},
\ref{rho_disc}).
The two lattices however have different symmetry and hence
 a {\it qualitatively} different spectroscopic behavior, as already
discussed in Sec. \ref{spectrum}.
 Lattice B has one molecule per unit cell and all
$k=0$ state are optically allowed, even if most of them have 
negligible intensity. The two molecules per cell  in lattice C are 
exchanged by reflection, so that  only
$k=0$ antisymmetric states are accessible by one-photon
absorption from the (totally-symmetric) gs: about half of the exact
$k=0$ states in Fig. \ref{ecci_disc_c}  are actually
 dark states. The special low-energy state with  large $n$
is a dark state in C-clusters, but corresponds to the state with
the largest  oscillator strength in B-clusters (cf Fig. 8, and 9).
The spectroscopic behavior of B-lattices is therefore particularly 
interesting and will be discussed in greater detail below.

Data in Fig. \ref{ecci_disc_b} show  that upon absorption of a single photon
about 3 excitons are created in B-lattice, or $\sim$ 4 molecules are
switched from N to I. This implies of course that the motion of 
 $\sim 4$ electrons is driven by the absorption of a single photon.
In B-lattices near the discontinuous interface 
the primary photoexcitation event therefore corresponds to a  
{\it multielectron transfer}.\cite{jacs} This sharply contrasts with 
the excitonic picture where  a single molecule 
 is excited by a single photon, 
leading to the transfer of no more than a single electron 
(specifically  $|1-2\rho| \le 1$ 
electrons are transferred upon photoexcitation).
Multielectron transfer is  a  new phenomenon with no
counterpart in the standard description of optical excitations in MM.
The theoretical and practical implications of multielectron transfer 
are hardly overemphasized.\cite{multi} Here we underline that so far 
multielectron transfer was discussed as a secondary photoexcitation 
 event: the absorption of a photon induces a single electron transfer, then
cooperative interaction with slow degrees of freedom (either vibrational
or environmental) leads to multiple electron transfer via a cascading
effect.\cite{multi} In our model direct (vertical)
photoexcitation  of several electrons is made possible by the 
correlation of electronic motion on different (non-overlapping)
molecules for systems at the discontinuous charge crossover.\cite{jacs}

\begin{figure}
\begin{center}
\includegraphics* [scale=0.5] {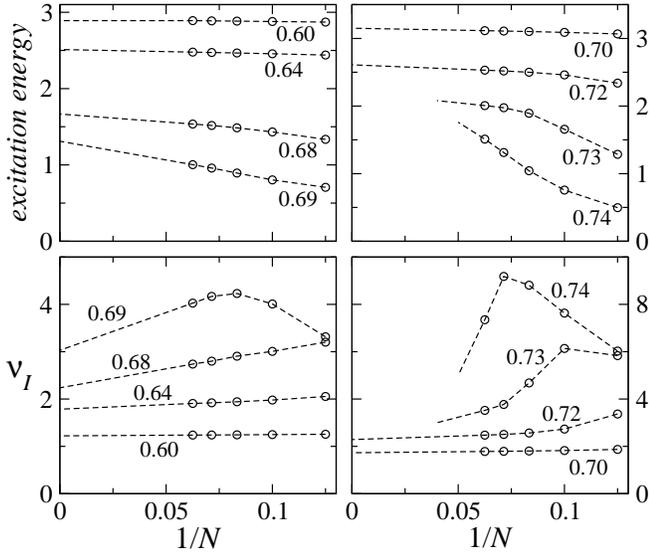}
\end{center}
\caption{The $1/N$ dependence of the 
excitation energy and of the average number of molecules turned I
 upon photoexcitation. Results refer to the state with the largest
transition dipole moment for  B-clusters with $v=2$, and
$z_0$=1.5, and 2 (left and right panels, respectively). 
The numbers labelling each curve show the corresponding $w$ value. 
 Lines are drawn as guide for eyes.}
\label{estrapola1}
\end{figure}

Data in Fig. \ref{ecci_disc_b} refer to a 10 site cluster, 
but multielectron transfer 
safely  survives in longer chains. With reference to the excited state
with the largest transition dipole moment, 
we calculate both the excitation energy and the average number of molecular
sites turned I upon excitation ($\nu_I$, defined as the difference 
between $n_I$ in the excited and ground state). Left panels in  
 Fig. \ref{estrapola1} show the N-dependence of these two
quantities calculated for a B-lattice with $v=2$, $z_0=1.5$, and several $w$.
For $w$ increasing from 0.60 to 0.69 the system is driven towards the interface
(cf Fig. \ref{rho_disc}) and $\nu_I$ smoothly extrapolates 
to $N\rightarrow \infty$,
with values that increase from $\sim$ 1.2 ($w=0.60$) to 2.5 ($w$=0.68).
Finite-size effects are larger at $w$=0.69, but $\nu_I$ clearly
extrapolates to even larger ($\sim$3) values there. 
The corresponding excitation energy smoothly extrapolates 
towards finite values in the $N\rightarrow \infty$ limit 
(Fig. \ref{estrapola1}, upper panel): in an infinite linear chain with 
B-geometry,  near the discontinuous
charge crossover the absorption of a single optical photon 
(with energies $\sim $ 1 eV for the parameters in Fig. \ref{estrapola1} and
$\sqrt{2}t$=1eV) drives the coherent motion
of several ($\sim$ 3) electrons.

The right panels in Fig. \ref{estrapola1} show the behavior of
a B-lattice near to a  discontinuous charge crossover with a much
larger ionicity jump with respect to case shown in the left panel
(cf. Fig. \ref{rho_disc}). Results are even more impressive there:
depending on the model parameters, in fact up to 8 electrons can be transferred
by absorption of a single photon. However, the finite-size analysis 
 requires in this case much longer chains than addressed here
(at least a few times longer than the number of electrons transferred).

To better understand the physics of multielectron transfer
we define the following $l$-th order correlation function:

\begin{equation}
f_l=\sum_{i=1}^N \langle \hat \rho_i \hat \rho_{i+1}...\rho_{i+l-1}
\rangle -\sum_{i=1}^N \langle \hat \rho_i\rangle ^l
\label{fofn}
\end{equation}
 $f_l$ vanishes exactly
for uncorrelated states,  that is for states that  can be reduced to the
 product of local molecular states. Positive (negative) $f_l$ indicates an
increased (decreased) probability of finding $l$ nearby I molecules
with respect to the uncorrelated state with the same average polarity.
\begin{figure}
\begin{center}
\includegraphics* [scale=0.6] {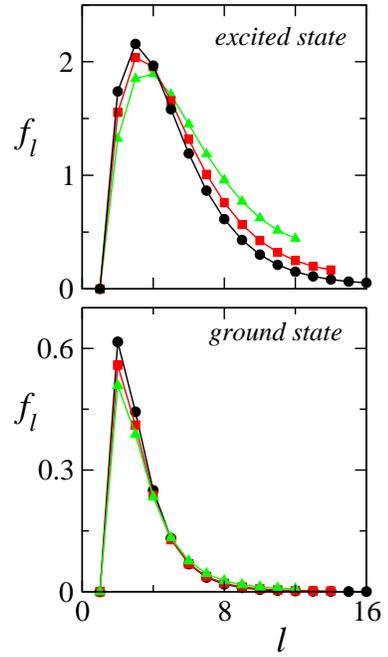}
\end{center}
\caption{The $l$ dependence of $f_l$ correlation function
calculated for the ground state and for the excited state with the
largest transition dipole moment for a B-lattice
with $v=2$, $z_0=1.5$ and $w=0.69$. Triangles, squares and circles
shows results for $N$= 12, 14, and 16, respectively. }
\label{corre}
\end{figure}
The upper panel of Fig. \ref{corre} shows the $l$-dependence of $f_l$
calculated for the most optically active state of a B-cluster with the same
parameters as in Fig. \ref{estrapola1}, just at the N-I crossover ($w=0.69$). 
In this state a non-negligible weight  is found of wavefunctions 
with several (2-4) nearby fully-I molecules. 
Then the photo-induced multielectron transfer corresponds to a {\it concerted 
electronic motion occurring on several
nearby molecules}.

The bottom panel in Fig. \ref{corre} shows the correlation function calculated
for the gs: quite predictably the overall degree of correlation is
smaller in the ground than in the excited state (notice the
different scale on the y-axis in the two panels in Fig. \ref{corre}). 
In any case data in Fig. \ref{corre} demonstrate a finite
contribution to the gs of I-droplet states, i.e. of
states  with two or
more nearby I-sites. Near the discontinuous charge crossover  
the mf description of  the gs fails: the exact gs is not the vacuum state
(cf the finite $n_e$ relevant to the exact gs in Fig. \ref{ecci_disc_b})
and cannot be described as the direct
product of local molecular states (cf the finite $f_l$ in the lowest 
panel in fig. \ref{corre}). Within the proposed model for {\it non-overlapping}
molecules {\it  classical}
electrostatic interactions lead, near the discontinuous crossover, to 
 an intrinsically {\it collective and correlated} zero-temperature gs.
The finite amplitude of I-droplet states in this gs is the key to understand
multielectron transfer: in B-clusters, I droplet states have a finite
(and large) permanent dipole moment: their finite amplitude in the gs is
therefore responsible for the appearance of sizeable transition dipole moments
from the gs towards states with a large I-droplet character. 
This observation also  sheds light on the qualitatively different behavior
of C and B lattices. For sure the discontinuous crossover in C-lattices
is similar in nature to that observed in B-lattices and correlated ground
and excited states are expected for these lattices too. 
However C-lattices are centrosymmetric: droplet states
 do not have permanent dipole moments and do not  contribute to 
the transition dipole moments.

The behavior  of systems lying near the discontinuous crossover, but on the I
side is similar to that of N systems, described above, provided the 
role of $|DA\rangle$ and $|D^+A^-\rangle$  is interchanged.
Similar behavior as described in Figs. \ref{estrapola1}, and 
\ref{corre} can be obtained in fact if
$\nu_I$, the number of sites turned I upon photoexcitation is substituted
by $\nu_N$, the  number of sites turned N.
Similarly $1-\hat\rho_i$ must substitute $\hat\rho_i$ to define a correlation 
function measuring  the
probability, in each relevant state, of droplets of $l$ nearby
N-molecules.

\section{\label{discussion}discussion}

In this paper we have presented a simple and interesting model for
clusters of polar and polarizable molecules. Each molecular site
is described in terms of a two-state model. Only classical electrostatic
 intermolecular interactions are accounted for, neglecting any intermolecular
overlap of wavefunctions on different molecules. 
The model applies quite naturally to MM 
(crystals, aggregates, films..) based on push-pull chromophores, an interesting
class of molecules quite extensively investigated for NLO applications.

Push-pull chromophores are largely polar and polarizable molecules,
and electrostatic interactions  can only
be understood within models that  properly
account for the molecular polarizability
at all orders: each molecule in fact experiences in the material 
the local electric fields generated by the surrounding molecules in 
 a non-trivial feed-back mechanism that is responsible for large collective
and cooperative effects.
The Mulliken model, adopted for  the molecular sites,
is particularly interesting in this respect since 
it accounts for the molecular  polarizability and 
hyperpolarizabilities,\cite{apcpl}
in a two-state approach that represents the simplest 
picture to describe polar molecules. 

Our model of course maps exactly onto the Hamiltonian derived several 
years ago by Agranovich for a molecular crystal described as a 
collection of two-state molecules interacting via electrostatic 
forces.\cite{agranovich} 
Being interested in the very weak interaction 
limit, Agranovich defined the excitonic Hamiltonian 
by disregarding all terms in the complete Hamiltonian not conserving the 
exciton number.\cite{agranovich} 
For polarizable molecules we propose a different strategy:
via a rotation of the local (on-site) basis we  retain the local 
term not conserving $n$. In other words we recognize that the optimal local
basis for the excitonic problem coincides with the eigenstates of the
local self-consistent mf problem. The relation between the mf and the 
excitonic approach is fairly fundamental. In the excitonic model
excitations are created on top of an uncorrelated gs: the mf solution 
then gives the best gs for the excitonic problem. 
The definition of the local states for the excitonic
problem as the eigenstates of the local
mf Hamiltonian unambiguously
defines  the excitonic and ultraexcitonic  Hamiltonian for any 
supramolecular arrangement, based on  the adopted model for 
the isolated molecule.
Uncovering a direct link from the molecular to the supramolecular 
description  gives  an important contribution to our understanding of MM,
and is a fundamental step to devise approaches to guide the chemical 
synthesis of functional  MM from the molecular to the
supramolecular level.

The mf description of MM can be set up for different
kinds of molecules described at different level of sophistication.\cite{ts} 
So our approach  is widely applicable.
 Here we have discussed its application to a simple model for
clusters of polar and polarizable chromophores where particularly important
collective and cooperative effects are expected. 
Collective effects show up in the gs with the large dependence of the
molecular polarity on supramolecular interactions. 
The effect of the surrounding medium, and specifically of the solvent,
 in tuning
the polarity of push-pull chromophores has already been 
underlined,\cite{reichardt,chemphys} and is a 
natural consequence of the large molecular polarizability. Here we extend this
concept to supramolecular interactions in MM and demonstrate that 
they can widely tune the molecular polarity, and eventually drive it 
across the N-I interface. The major difference with 
respect to solvated molecules is the possibility for MM 
to support true phase transitions, as the extreme 
manifestation of cooperative behavior.

The properties of isolated 
push-pull chromophores, and most interestingly, their NLO
responses strongly depend on the molecular polarity:\cite{apcpl}
  the large variation of $\rho$ induced 
by supramolecular interactions then immediately suggests that NLO responses
are strongly  affected by supramolecular interactions. 
Apart from a narrow region around the crossover 
regime, the mf approach  nicely approximate the  NLO
responses of the material. Huge collective effects are demonstrated 
that are only partly accounted for within the excitonic approximation.
The large ultraexcitonic corrections in Fig. \ref{nlo_ff}
suggest that the 
excitonic model does not provide a reliable approximation scheme
 for the calculation of 
NLO responses of MM based on largely polarizable molecules.

The traditional  strategy to get 
molecular materials with large $\beta$ responses relies on the
optimization of $\beta$ at the molecular level, in the 
tacit assumption that materials based on molecules with the largest $\beta$
will have the largest responses.
  But, in agreement with experimental observation,
\cite{mukamel,dalton,ciclo} results 
 in Fig. \ref{nlo_ff} demonstrate that
 the properties of MM based on push-pull chromophores 
are far from additive. 
To guide the synthesis of optimized materials,
structure properties relationships must be devised
and fully understood not only at the molecular level but also
at the supramolecular level. The detailed analysis of the simple model for
supramolecular clusters  presented here is  a first step in this direction.

Deviations of the excitation spectrum from the excitonic description are 
predictably sizeable for materials with medium-large supramolecular 
interactions, with more important effects for materials located near the
N-I interface where the molecular polarizability is at a maximum. 
More interesting is the observation of new phenomena driven by cooperativity
near the discontinuous charge-crossover of attractive lattices. 
The lowest excitation of a N (I) lattice near the N-I 
discontinuous crossover creates an droplet of I (N) molecules whose
size increases as the system is driven towards more discontinuous  interfaces.
These droplet states, that appear as low-lying excitations near the 
discontinuous N-I crossover, share the same physics 
with the charge-density-wave 
droplets observed in half-filled
extended Hubbard 1D chains at  the discontinuous phase transition from the 
 Mott-insulator to the charge-density wave state,\cite{je}
and are due to the nucleation of I (N) domains
in the presence of large intersite interactions.
The main novelty of our result in this respect is
the observation that these droplet states are, in a non-centrosymmetric
environment, optically allowed, so that the absorption of a single 
photon in a N (I) material can directly create an I (N) droplet extending over
several molecules. The primary photoexcitation event is in these
conditions a multielectron transfer that is a direct consequence of
the correlation of the electronic  motion  among different 
non-overlapping molecules.

We have discussed the gs and the excitation spectrum of a model for polar and
polarizable molecular units, only interacting via electrostatic forces. 
The model naturally applies to MM based on push-pull chromophores, 
but also offers a rough description of organic CT salts with a mixed stack
motif. In these materials, $\pi$-conjugated molecules with a strong
electron  D and A character  alternate  in a one-dimensional stack:
the sizeable overlap between adjacent molecules leads to charge-resonance
and hence to fractional charges on molecular sites. The N-I 
phase transition observed in these materials\cite{nitearly}
 is a fascinating phenomenon whose physics is actively 
investigated.\cite{tokura}
 In the early days of the N-I transition, it was modeled by 
reducing the  stack of overlapping molecules into a 
collection of non-overlapping DA pairs,\cite{soos}
 i.e. by adopting exactly the same model proposed here for B-lattice. 
Of course this approximation is very rough and more refined models 
are needed to understand the complex and variegated behavior of  CT
salts with a mixed stack motif.\cite{anu}  
However, the N to I crossover discussed here for attractive lattices 
of polar and polarizable molecules
shares the same physics as the N-I transition in CT salts. 
In this respect we underline that anomalous spectral features for CT salts
in the region of the discontinuous N-I transition were already 
discussed in Ref.\onlinecite{soosopt}. 
In particular the importance of I/N droplet 
states as well as the failure of the excitonic approximation in the proximity
of the discontinuous crossover were addressed. 
Multielectron transfer, described here for molecular 
clusters, indeed offers suggestive hints on the physics of photoinduced
N-I phase transitions in CT salts.\cite{jacs}
However CT salts are extended systems, with electrons delocalized all
along the stack. On the 
opposite, the molecular clusters described here are fully localized systems:
electrons cannot escape from their parent molecule. 
In this respect we  underline that the main novelty and interest of the present
work is in the observation of 
 cooperative phenomena and correlation effects among {\it non-overlapping}
molecular units.

\begin{acknowledgments}
A.P. thanks  Z.G.Soos for several enlightening discussions, 
and S.Mazumdar, S.Ramasesha and F.Spano for
useful conversations. F.T. thanks J.Knoester and V.Chernyak 
for interesting discussions. 
Work partly supported by the Italian Ministry of Education (MIUR) through 
COFIN-2001, and by INSTM through PRISMA-2002.
\end{acknowledgments}

 \end{document}